\documentclass[proof]{WileyASNA-v1}

\articletype{Article Type}%

\usepackage{subfig}
\received{22 September 2019}
\revised{11 November 2019}
\accepted{23 January 2020}

\begin{document}

\title{SPI Analysis and Abundance Calculations of DEM L71, and Comparison to SN explosion Models\protect}

\author[1]{Jared Siegel}

\author[1]{Vikram V. Dwarkadas}

\author[2]{Kari Frank}

\author[3] {David N. Burrows}

\author [1] {Aldo Panfichi}

%\authormark{AUTHOR ONE \textsc{et al}}

\address[1]{\orgdiv{Astronomy and Astrophysics}, \orgname{University of Chicago}, \orgaddress{\state{Chicago, IL}, \country{USA}}}

\address[2]{\orgdiv{CIERA}, \orgname{Northwestern University}, \orgaddress{\state{Evanston, IL}, \country{USA}}}

\address[3]{\orgdiv{Astronomy and Astrophysics}, \orgname{Penn State University}, \orgaddress{\state{University Park, PA}, \country{USA}}}

\corres{Vikram Dwarkadas. \email{vikram@astro.uchicago.edu}}

\presentaddress{Department of Astronomy and Astrophysics, University of Chicago, 5640 S Ellis Ave, ERC 569, Chicago, IL, 60637}

\abstract{We analyze the X-Ray emission from the supernova remnant DEM L71 using the Smoothed Particle Inference (SPI) technique. The high Fe abundance found appears to confirm the Type Ia origin. Our method allows us to separate the material ejected in the supernova explosion from the material swept-up by the supernova shock wave.  We are able to calculate the total mass of this swept-up material to be about 228 $\pm$ 23 M$_{\odot}$. We plot the posterior distribution for the number density parameter, and create a map of the density structure within the remnant. While the observed density shows substantial variations, we find our results are generally consistent with a two-dimensional hydrodynamical model of the remnant that we have run. Assuming the ejected material arises from a Type Ia explosion, with no hydrogen present,  we use the predicted yields from Type Ia models available in the literature to characterize the emitting gas. We find that the abundance of various elements match those predicted by  deflagration to detonation transition (DDT) models. Our results, compatible with the Type Ia scenario, highlight the complexity of the remnant and the nature of the surrounding medium.}


\maketitle

\section{Introduction}
\label{sec:intro}

\subsection{SPI} Supernova remnants (SNRs) are complex, three-dimensional objects; 
properly accounting for this complexity in the resulting X-ray
emission presents quite a challenge. Smoothed
Particle Inference \citep[SPI,][]{spi} is a flexible technique for
fitting X-ray observations of extended objects developed specifically
to address this problem. It owes its flexibility to its modeling of
the plasma as a collection of independent `smoothed particles,' or
blobs, of plasma. Each blob can have its own model parameters,
including temperature, abundance, spatial position, and size. It is not necessary to assume any particular morphology or symmetry,
and a multiphase plasma can be modeled using multiple
independent blobs. SPI uses a Markov Chain Monte Carlo approach to
iterate over the blob model parameters, forward folding the blob model
through the {\it XMM-Newton} instrument response and comparing to the
data. The distributions of any number
of plasma properties can be characterized from the posterior distributions of the relevant
blob parameters.

\subsection{DEM L71} DEM L71 is an $\approx$4000 year old supernova remnant in the LMC. 
It has a more or less regular shape, and has been classified
previously as a Type Ia SNR by several authors, based on excess Fe
abundance in the central part of the remnant \citep{hughesetal03,
  rakowskietal03, vanderheydenetal03, ghavamianetal03}

\begin{figure}[!tbp]
  \centering
  \subfloat[]{\includegraphics[width=0.39\columnwidth]{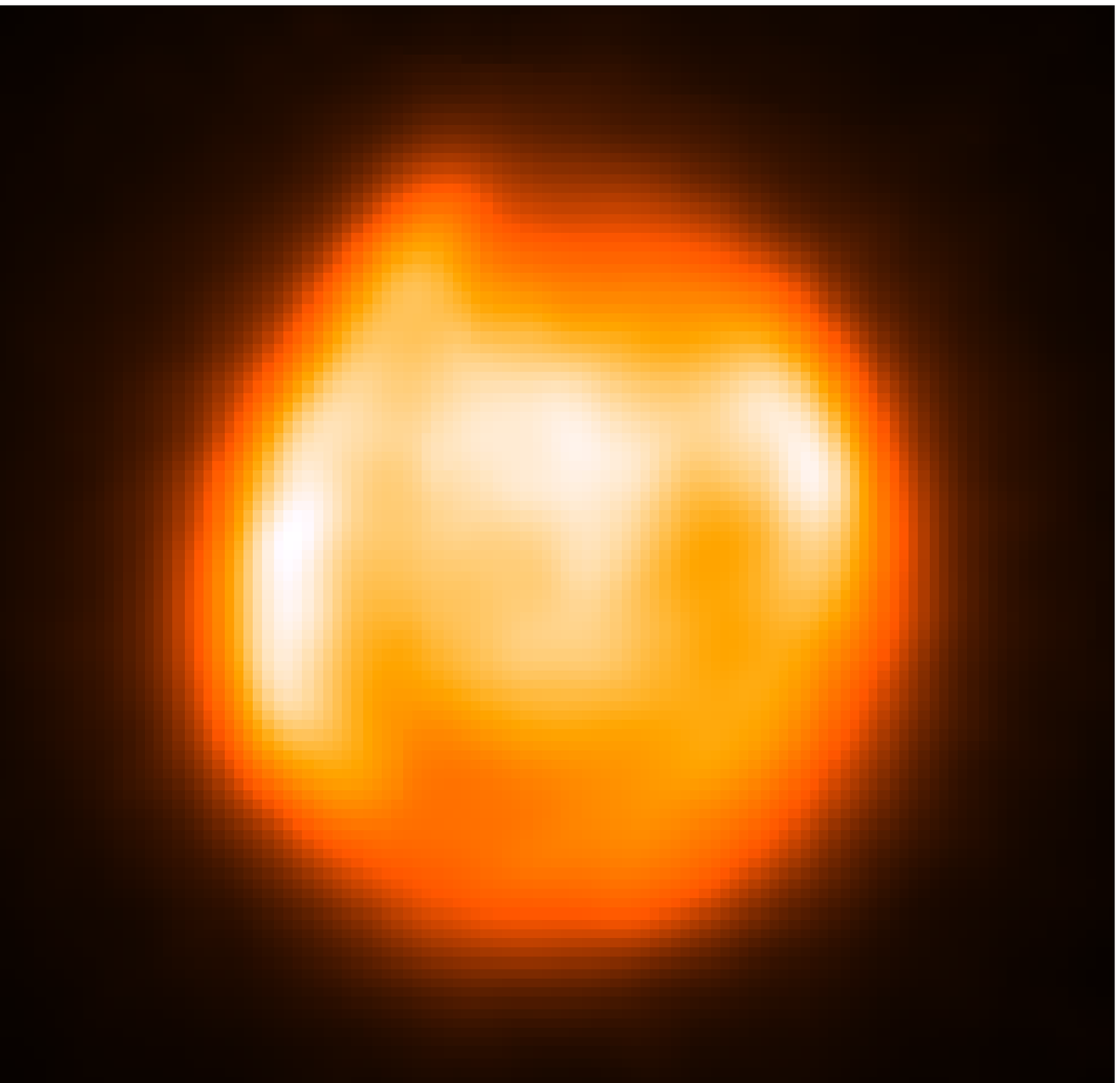}}
  %\hfill
  \subfloat[]{\includegraphics[width=0.51\columnwidth]{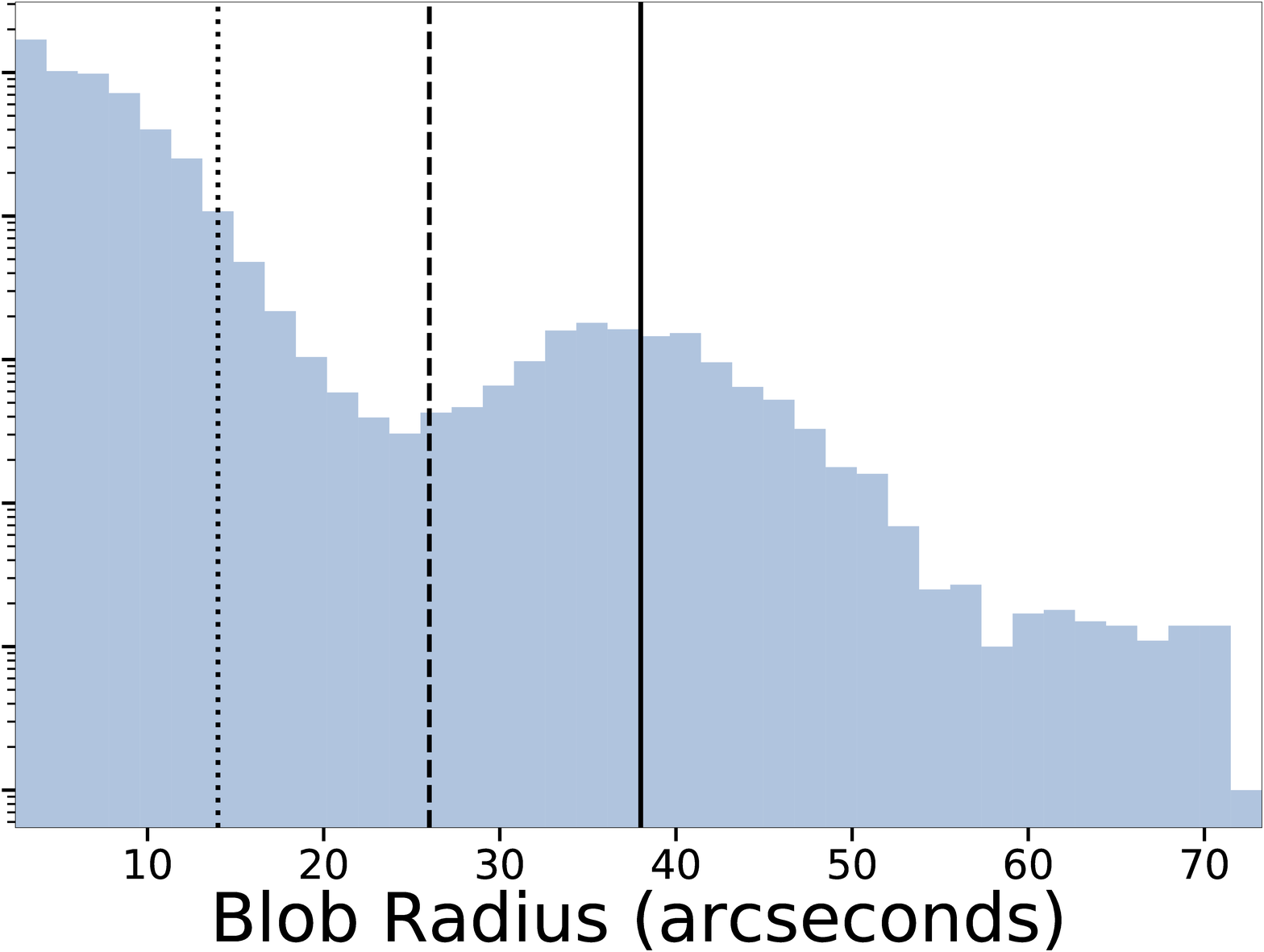}}
  \caption{Left: Combined EPIC-pn and MOS exposure-corrected image in the 0.2-8.0 keV band. Right: EM-weighted blob radius histogram for the entire remnant, with the radius of the entire remnant (solid line), the central emission region (dashed line), and the width of outer shell (dotted line) also shown.}
  \label{fig:imageblobsize}
\end{figure}

In \citet{franketal19} we applied the SPI technique to XMM EPIC observation 0201840101 of DEM L71 from
2003 December (see Figure \ref{fig:imageblobsize}(left) for associated image). We used an absorbed vpshock model to fit the SNR emission, along with
several components to account for the different types of X-ray
background emission. Each iteration of the SPI fitting process used fifty blobs and blobs from all converged iterations were used in the final analysis. A histogram of all blob radii is shown in Figure \ref{fig:imageblobsize} (right). The temperature, ionization age, and abundance of
O, Ne, Mg, Si, S, and Fe were thawed, and independent for each
blob.  Here we extend the analysis of the SPI fit by
calculating the composition of the swept-up material and the ejecta of
DEM L71, and comparing those to a large set of supernova explosion
models.

\section{Methods}
\label{section:methods}

\subsection{Separating Emission from Ejecta and Swept-Up Material}
\label{section:separating}

\citet{franketal19} included maps showing the distribution  of O, Ne, and Fe.The maps showed that O and Ne abundances are low throughout the entire remnant, while Fe is similarly low in the outer shell but significantly enhanced in the central emission region.
The central region was therefore assumed to signify the ejected material from the explosion, which was isolated by selecting
blobs with parameters kT $>1$ keV, Fe/Fe$_{\odot}>$1.  Here we apply
the additional criterion that the blob radius be less than 26", the
approximate radius of the central emission, see Figure \ref{fig:imageblobsize}. These larger blobs
comprise only 3.5\% of all the blobs and a small percentage of total
emission measure (EM) of the kT $>1$ keV, Fe/Fe$_{\odot}>$1 blobs. We
have confirmed that the large blobs are unlikely to represent the
central ejecta. However they have relatively large mass due to their
large volumes, and thus disproportionately bias any mass estimates of
the ejecta. All blobs not designated as ejecta are considered to be
swept-up material. The total EM of the ejecta
is EM = 8.82 ($\pm$ 2.31) $\times$ 10$^{57}$ cm$^{-3}$, and that of
the swept-up material is 5.44 ($\pm$ 0.59) $\times$ 10$^{59}$
cm$^{-3}$.

\subsection{Mass and Density}
\label{section:masscalc}

The density of a blob can be calculated from its emission measure and
volume:
\begin{equation}
\label{eqn:EM}
    EM = \int{n_en_HdV}
\end{equation}
\noindent
where $n_e$ is the electron density and $n_H$ is the hydrogen
density. The EM is obtained directly from the vpshock model
normalization. The blob radius, and thus volume (assumed to have a Gaussian profile), is a free parameter
in the SPI fit, and can be computed for each blob. The mass can be
derived from the density and the volume. However, equation
\ref{eqn:EM} provides only the product of the electron and hydrogen
densities. One more equation involving the two quantities is needed in order to obtain them individually.
We consider three possible scenarios via which we can calculate these quantities. In each scenario, we assume that a blob has a
uniform density, and the plasma is fully ionized. We have
confirmed that assuming a partially ionized plasma does not appear to
have a significant effect on our results, and any modifications
generally lie within the error bars. The mass of each blob is derived
individually, and summed over all blobs to determine the total mass.

The simplest scenario is to assume that all the emitting material
consists of `typical' LMC plasma, and to consider the EM and volume
from the SPI fit. In this case, we use `typical' abundances for the
LMC as listed in \citet{rd92}. This leads to an $n_e/n_H$ ratio of
1.087 and $\mu=0.602$ for each blob.

The second scenario allows for fit elements to deviate from the `typical' LMC values.
For those
species thawed in the vpshock model, the abundance values are taken
from the SPI fit, while the remaining species are assumed to have a
`typical' LMC abundance, as defined in \citet{rd92}. Depending on the location of a blob, various different components, such as ejecta, swept-up medium or local LMC material, may contribute to the abundance value.
Since each blob has a unique abundance value calculated from the SPI
fit, it will have a unique value of $n_e/n_H$ and $\mu$. The mass of
each blob is computed individually, and then they are all summed.

In these two scenarios, we calculate the mass of an individual element using the element's mass fraction in the remnant. This fraction is calculated using the ratios $N_i/N_{ions}$, the number of atoms of a
given element over the total number of atoms, and $m_i/\bar{m}$, the
molecular weight of the element over the average ion mass. From this mass
fraction and the total mass, the individual mass of a given element
present in the plasma can be calculated.

The third scenario assumes the ejecta arise from a Type Ia explosion, where  no hydrogen is present. It is therefore inappropriate to
use the abundance returned by the SPI fit, because XSPEC 
assumes the presence of hydrogen. In order to compute the
ejecta mass, we instead choose a Type Ia explosion model available in
the literature, and use the given yields to specify the
composition. Any Type Ia model that provides the mass yield per
element can be used. The models we use are listed in Table
\ref{table:models}. Given the yield of each element, and assuming it
to be fully ionized, we can compute the number of free electrons, and
the mean molecular weight $\mu$, in the same manner as the other
scenarios. The main difference here is that we are assuming the
composition is {\em not} defined by the SPI fit but by the Type Ia
model. We note that we adopt the mass fraction of each element as
given in a particular Type Ia model, but not the total mass predicted by
the model. The total mass will be derived from the SPI volume and the
density estimated. In some cases this may overestimate the mass. This
may be because there is some swept-up material mixed in with the
ejecta, but a large overestimate would generally mean that the assumption of a Ia is either untrue, or that even if it is true, the selected model is a poor approximation for the emitting gas.

\section{Results and Discussion}
\label{results}

\subsection{Entire Remnant}
\label{section:entireremnant}

\begin{figure}[h]
	\centerline{\includegraphics[width=78mm]{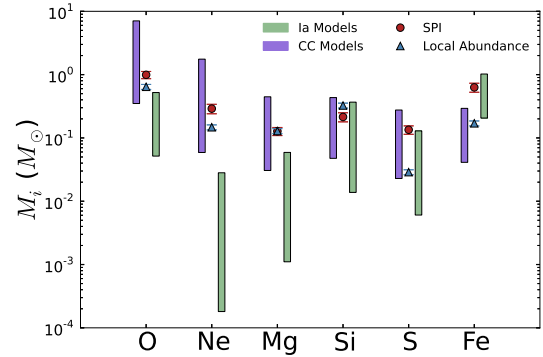}}
	\caption{The mass of each element over the entire remnant,
          derived using the SPI measured abundances (\textit{red
            circle}) and the Typical LMC scenario (\textit{blue
            triangle}), in comparison with the range of values
          predicted by the core collapse (\textit{purple bar}), and
          Type Ia (\textit{green bar}) models listed in Table
          \ref{table:models}. All masses are in units of solar
          mass.\label{fig:allyields}}
\end{figure}

We first compare the total mass of each element for all the blobs with a selection of core collapse and Type Ia models. For each element, the mass is inferred from all the blobs using both the SPI measured abundance and the `typical LMC'
scenario. These are shown in Figure \ref{fig:allyields}, alongside the
range of masses predicted by both core collapse and Type Ia explosion
models. The models considered are listed in Table
\ref{table:models}. The measured masses of O, Ne, and Mg appear to
correspond well with the core collapse models. However, these
abundance values are also consistent with what would be expected from
swept-up ambient LMC material. Although they seem a bit higher, this
may be because the `typical LMC' values in reality have considerable
variation depending on position, latitude and the stellar environment,
that is not accounted for. Alternatively, it is possible that these
values may be fit by the typical LMC value with some contribution from
the Type Ia material. Therefore the small enhancement of O, Ne, and Mg
masses above the LMC value is not indicative of a core-collapse (CC)
origin.

The predicted mass range for Si and S overlaps both the CC and Type Ia
models, and has little discriminating power. The high measured Fe
abundance, however, is totally inconsistent with an origin either in
typical LMC material, or in core-collapse explosions. It can only be
matched by the mass range predicted by Type Ia models. The Fe
abundance therefore has the most discriminating power, and clearly
suggests a Type Ia rather than a core-collapse SN explosion. In this
we are consistent with previous authors, although with SPI we have the
ability to determine abundances and other properties throughout the
remnant as well as in any individual location.

\begin{center}
\begin{table}[t]%
\centering
\caption{Supernova Explosion Models Used for Comparison.\label{table:models}}%
\tabcolsep=0pt%
\begin{tabular*}{20pc}{@{\extracolsep\fill}lcccc@{\extracolsep\fill}}

\toprule
\textbf{Model Name} & \textbf{Group}  & \textbf{Reference}\\
\midrule
N10  & Ia DDT  & S13  \\
N100  & Ia DDT  & S13  \\
N1600  & Ia DDT  & S13  \\
N1600 0.5Z  & Ia DDT  & S13  \\
050-1-c3-1  & Ia DDT  & L18  \\
300-1-c3-1  & Ia DDT  & L18  \\
500-1-c3-1  & Ia DDT  & L18 \\
C-DEF  & Ia DEF  & M10  \\
050-1-c3-1P  & Ia DEF  & L18  \\
300-1-c3-1P  & Ia DEF  & L18  \\
500-1-c3-1P  & Ia DEF  & L18  \\
W18 s12.5 & CC  & S16  \\
W18 s18.1 & CC  & S16  \\
W18 s25.2 & CC  & S16  \\
W18 s60.0 & CC  & S16  \\
20M$_{\odot}$ 10E$_{51}$ & CC & N06  \\
40M$_{\odot}$ 30E$_{51}$ & CC & N06  \\
25A & CC & M03  \\
25B & CC & M03  \\
40A & CC & M03  \\
40B & CC & M03  \\
\bottomrule
\end{tabular*}
\begin{tablenotes}
\item S13:\citet{seitenzahletal13}, L18:\citet{ln18}, M10:\citet{maedaetal10}, S16:\citet{sukhboldetal16}, N06:\citet{nomotoetal06}, M03:\citet{mn03}

\end{tablenotes}
\end{table}
\end{center}

\subsection{Swept-Up Material}
\label{section:swept}

\begin{figure}[htbp]
	\centerline{\includegraphics[width=78mm]{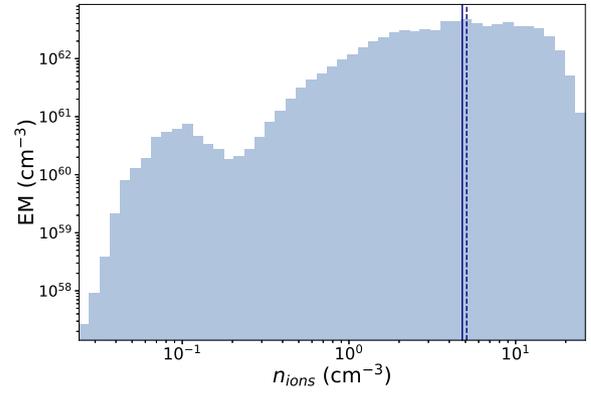}}
	\caption{ EM-weighted density histogram for the swept-up
          material, using the SPI measured abundances for the
          composition. The mode and median are shown as dashed and
          solid lines, respectively.
	\label{fig:sweptdensitydist}}
\end{figure}

\begin{figure}[htbp]
	\centerline{\includegraphics[width=78mm]{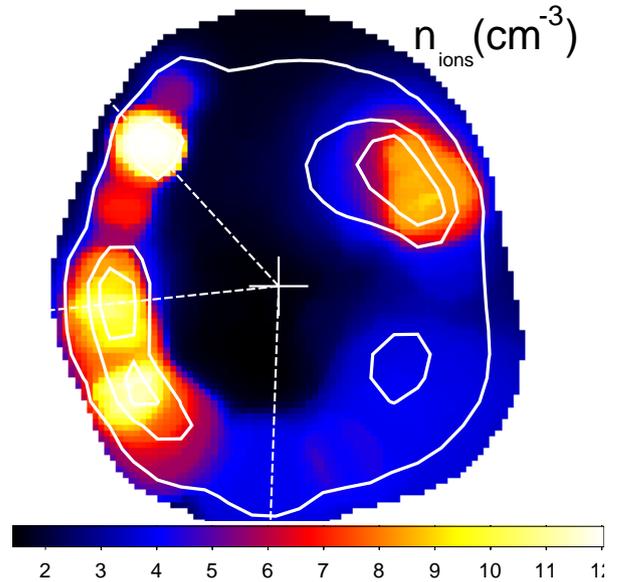}}
	\caption{EM-weighted density map of the swept-up medium, with
          the density calculated using the SPI measured abundances. The
          white contours are from the overall EM map. The cross indicates the geometric center used
          to create the density profiles in Figure
          \ref{fig:sweptdensityprofile}. Density is set to zero where
          the EM was lower than 0.05\% of the mean to avoid noise due
          to poor statistics on the outer
          edges. \label{fig:sweptdensitymap}}
\end{figure}

The density histogram of the swept-up material is shown in Figure
\ref{fig:sweptdensitydist}. The distribution results in an EM-weighted median
density of $4.8\pm1.1$ cm$^{-3}$. A map of the density is shown in
Figure \ref{fig:sweptdensitymap}. The densities derived 
are found to be consistent with the densities from 
\citet{vanderheydenetal03}.

\begin{figure}[htbp]
	\centerline{\includegraphics[width=78mm]{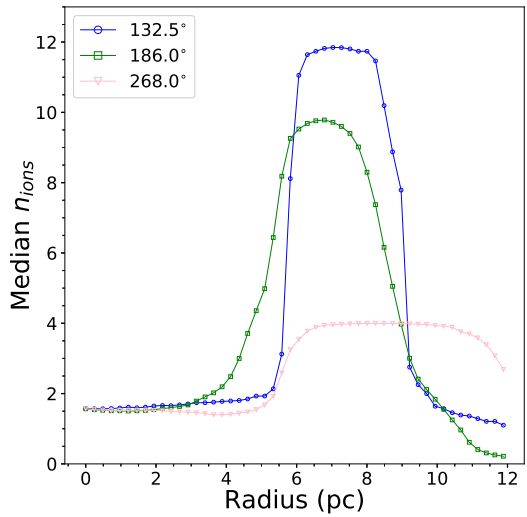}}
	\caption{EM-weighted density as a function of radius for the
          swept-up material for three different position
          angles. \label{fig:sweptdensityprofile}}
\end{figure}

Using the density map, we plot the EM-weighted median density versus
radius profile for three azimuthal positions (Figure
\ref{fig:sweptdensityprofile}). The selected position angles are shown
in Figure \ref{fig:sweptdensitymap} as white dashed lines. We 
clearly see  high density blobs in the swept-up medium
whose density far exceeds the `average' density in the swept-up
medium. The regions of highest density in Figure \ref{fig:sweptdensitymap} appear to coincide with regions showing high H$\alpha$ emission \citep{R09}, thus further suggesting some form of density enhancement in this area.

\begin{figure}[htbp]
	\centerline{\includegraphics[width=90mm]{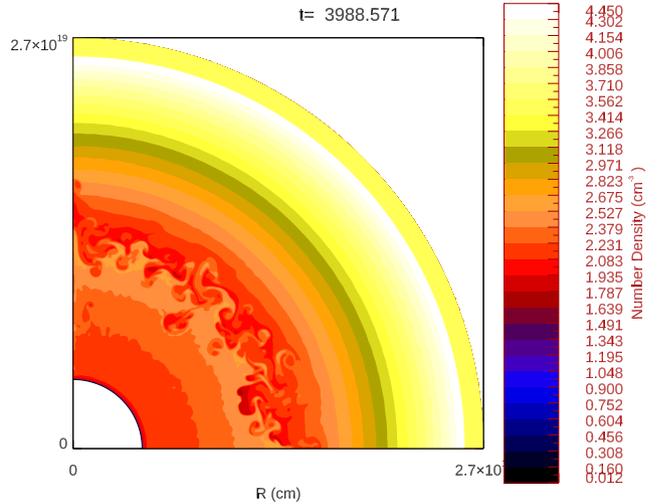}}
	\caption{ The number density of the plasma derived from our
          numerical simulations. The figure shows the density
          calculated by dividing the fluid density at each point by
          1. $\times 10^{-24}$, which essentially assumes a fully
          ionized medium. In theory this may not be valid at all
          points, and the mean molecular weight may be different
          within the ejecta than the surrounding medium, but within
          the approximations this will suffice.  \label{fig:densim}}
\end{figure}

As described in \citet{franketal19}, numerical hydrodynamic simulations
to compute the evolution of the remnant were carried out using the
VH-1 code, a 3-dimensional numerical hydrodynamics code based on the
Piecewise Parabolic Method \cite{cw84}. The simulations were run in
2-dimensions. One quadrant was simulated, assuming spherical
symmetry. As expected, the contact discontinuity between the inner and
outer shocks is unstable to the Rayleigh-Taylor (R-T) instability, and
R-T fingers are seen. Other than this there is nothing in the simulations that can break
the spherical symmetry. Thus for the most part the swept-up medium in
our simulation is quite uniform, as shown in Figure
\ref{fig:densim}. The density shown in this figure is the number
density, obtained simply by dividing the fluid density at each point
by 1. $\times$ 10$^{-24}$, which may be considered as assuming a mean
molecular weight $\approx$ 0.6, i.e. an assumption of complete
ionization. On the other hand, the density map obtained from SPI
(Figure \ref{fig:sweptdensitymap}), while also 2D, is in reality a 2D
projection of a 3D environment where material may be mixed from the
outset, and velocities in all directions may not necessarily be
uniform. The dense regions in the outskirts of this map may be due to
explosion inhomogeneities, an inhomogeneous surrounding medium, or
mixing in turbulent layers during the explosion, none of which are
captured in our simulations. Given this, it is comforting that the
number densities from the simulations and those calculated by SPI are
of the same order and differ by about a factor of 2 or so.

Since the swept-up material is assumed to have a `typical' LMC abundance of hydrogen, the density and mass of the swept-up material can be computed as defined in Section \ref{section:masscalc}. Using the measured SPI
abundance to determine the composition, we find a total mass of the
swept-up material of $228\pm23$ M$_{\odot}$. This mass is much larger
than that derived by \citet{vanderheydenetal03}. We suspect that this
could be due to the volume of our surrounding medium, which exceeds the
volume derived by \citet{vanderheydenetal03}. The latter used a
spherical shell defined by somewhat arbitrary boundaries. In our case
we isolated the ejecta as described above, and then computed the
volume of each blob (assumed to have a Gaussian profile) and added it
up to get the total volume. However we note also that we had some large volume blobs that we discarded as ejecta and are therefore now considered to be swept-up medium. Furthermore, the high density in the outer parts may be indicative of large density variations that were not considered by \citet{vanderheydenetal03}.

The mass of swept-up material can be independently estimated under the assumption
that the SN expanded in a constant density medium, and that the
supersonic shock wave swept up all the material in its path. In this
case, assuming the density of the surrounding medium to be
${\rho}_{am}$, the swept up mass simply corresponds to $4 \pi /3
R_{sh}^3 {\rho}_{am}$, where $R_{sh}$ is the radius of the outer
shock. The latter is essentially the radius of the remnant.  In our simulations we measured a shock radius of 8.5pc and a number density of 1.13 $\times 10^{-24}$ g cm$^{-3}$.  This gives a mass of swept-up material of 42 M$_{\odot}$. \citet{Leahyetal17} calculated a radius of 9.45 pc and a density of 1.28,  although they do not indicate what value of $\mu$ they use. For a value of 1, they would get about 110 M$_{\odot}$. A radius of 12 pc, close to the upper limit of suspected values, and closer to what we estimate from the data, would give a mass of 226 M$_{\odot}$ for the same value of $\mu$. Some of these values  seem lower than SPI gives. However it is worth keeping in mind the high density blobs that we see in the outer regions, with number density approaching 12, much higher than even the shock densities. This may indicate significant density variations in the surrounding medium, thus invalidating the assumption of a constant density medium, and leading to the higher mass we find for the swept-up material. 

\begin{figure}[htbp]
	\centerline{\includegraphics[width=78mm]{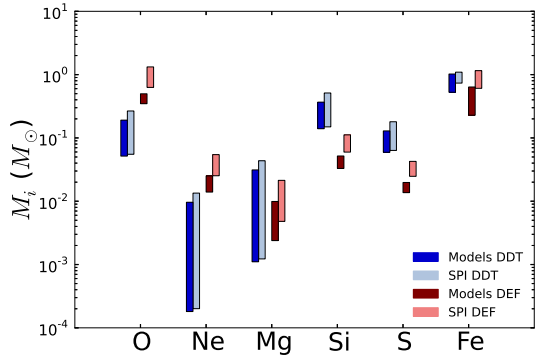}}
	\caption{The mass yield (in solar masses) predicted by deflagration-to-detonation (DDT) (\textit{dark blue bar}) and pure deflagration
          (\textit{dark red bar}) Type Ia models, as listed in Table
          \ref{table:models}, in comparison with the ejecta mass
          calculated from the SPI results by assuming pure Type Ia
          ejecta composition using DDT (\textit{light blue bar})
          and deflagration (\textit{light red bar}) models.  \label{fig:ejectamasses}}
\end{figure}

\subsection{Ejecta Mass} To derive the ejecta mass, we use the 
pure Type Ia scenario outlined in Section \ref{section:masscalc},
including only those blobs defined as
ejecta. We calculate the mass using each and every one of the
Type Ia models in Table \ref{table:models}. We consider the entire
mass range from all models.  The models are separated into two groups, those that consider a pure deflagration
(i.e. a subsonic burning front), and those that incorporate a
deflagration to detonation transition (i.e. a transition from a
subsonic to a supersonic front).  Our results are shown in Figure \ref{fig:ejectamasses}.  The
measured and predicted mass using the DDT models tend to agree very
well, consistent with results in general for Type
Ia's. However, the deflagration models are not consistent with
measurements.

\section{Conclusions}
\label{section:conclusions}

In this paper we have expanded on our previous work on SNR DEM L71
\citep{franketal19}.  We have better isolated the ejecta,
computed the abundance of various elements and compared to detailed SN
explosion models available in the literature. We have confirmed that
DEM L71 shows an excess of Fe in the central region, which
can only be produced within a Type Ia explosion, and is incompatible
with a core-collapse explosion. Furthermore the abundance values are more compatible with DDT models than pure deflagration models. Our simulations are reasonably consistent with our extracted density map, 
although the map reveals inhomogeneities in the density distribution that
cannot be captured in our simulations.

 We are now applying our SPI technique to other SNRs
observed with {\it XMM-Newton}, including W49B, and will show the results from these comparisons in future papers.

\section*{Acknowledgments}
This work was supported by NASA ADAP grant NNX15AH70G to Penn
State University, with subcontracts to the University of Chicago and
Northwestern University. Based on observations obtained with
XMM-Newton, an ESA science mission with instruments and contributions
directly funded by ESA Member States and NASA.

\subsection*{Author contributions}

Jared Siegel carried out much of this work under the supervision of, and directions from, Vikram Dwarkadas (on abundance calculations) and Kari
Frank (on SPI). Aldo Panfichi helped early on. David Burrows is in overall charge of the project and PI on the supporting grant.

\subsection*{Financial disclosure}

None reported.

\subsection*{Conflict of interest}

The authors declare no potential conflict of interests.

\bibliography{spipaper}%

\begin{thebibliography}{}

\bibitem [\protect \citeauthoryear {%
Bettonvil%
, Sutterlin%
, Hammerschlag%
, Rutten%
\BCBL {}\ \BBA {} Stix%
}{%
Bettonvil%
\ \protect \BOthers {.}}{%
{\protect \APACyear {2003}}%
}]{%
Bettonvil2003}
\APACinsertmetastar {%
Bettonvil2003}%
\begin{APACrefauthors}%
Bettonvil, F\BPBI C.%
, Sutterlin, P.%
, Hammerschlag, R\BPBI H.%
, Rutten, R\BPBI J.%
\BCBL {}\ \BBA {} Stix, M.%
\end{APACrefauthors}%
\unskip\
\newblock
\APACrefYearMonthDay{2003}{}{},
\newblock
{\BBOQ}\APACrefatitle {Proc. SPIE Conf. Ser.} {Proc. SPIE Conf. Ser.}{\BBCQ}
\newblock
\BIn{} \BVOL\ 4853, \BPG~306.
\PrintBackRefs{\CurrentBib}

\bibitem [\protect \citeauthoryear {%
Bland-Hawthorn%
, van Breugel%
\BCBL {}\ \BBA {} Gillingham%
}{%
Bland-Hawthorn%
\ \protect \BOthers {.}}{%
{\protect \APACyear {2001}}%
}]{%
Bland2001}
\APACinsertmetastar {%
Bland2001}%
\begin{APACrefauthors}%
Bland-Hawthorn, J.%
, van Breugel, W.%
\BCBL {}\ \BBA {} Gillingham, P\BPBI R.%
\end{APACrefauthors}%
\unskip\
\newblock
\APACrefYearMonthDay{2001}{}{},
\newblock
\unskip
\newblock
\APACjournalVolNumPages{ApJ}{563}{}{611}.
\PrintBackRefs{\CurrentBib}

\bibitem [\protect \citeauthoryear {%
Kosugi%
\ \BBA {} Gillingham%
}{%
Kosugi%
\ \BBA {} Gillingham%
}{%
{\protect \APACyear {2007}}%
}]{%
Kosugi2007}
\APACinsertmetastar {%
Kosugi2007}%
\begin{APACrefauthors}%
Kosugi, T.%
\BCBT {}\ \BBA {} Gillingham, R\BPBI H.%
\end{APACrefauthors}%
\unskip\
\newblock
\APACrefYearMonthDay{2007}{}{},
\newblock
\unskip
\newblock
\APACjournalVolNumPages{Sol. Phys.}{243}{}{3}.
\PrintBackRefs{\CurrentBib}

\bibitem [\protect \citeauthoryear {%
Kosugi%
\ \protect \BOthers {.}}{%
Kosugi%
\ \protect \BOthers {.}}{%
{\protect \APACyear {2009}}%
}]{%
Kosugi2009}
\APACinsertmetastar {%
Kosugi2009}%
\begin{APACrefauthors}%
Kosugi, T.%
, Matsuzaki, K.%
, Sakao, R.%
, Bettonvil, F\BPBI C.%
, Sutterlin, P.%
\BCBL {}\ \BBA {} Hammerschlag, R\BPBI H.%
\end{APACrefauthors}%
\unskip\
\newblock
\APACrefYearMonthDay{2009}{}{},
\newblock
\unskip
\newblock
\APACjournalVolNumPages{Sol. Phys.}{243}{}{3}.
\PrintBackRefs{\CurrentBib}

\bibitem [\protect \citeauthoryear {%
Power%
\ \protect \BOthers {.}}{%
Power%
\ \protect \BOthers {.}}{%
{\protect \APACyear {1975}}%
}]{%
Paivio1975}
\APACinsertmetastar {%
Paivio1975}%
\begin{APACrefauthors}%
Power, J\BPBI D.%
, Cohen, A\BPBI L.%
, Nelson, S\BPBI M.%
\ et al.\end{APACrefauthors}%
\unskip\
\newblock
\APACrefYearMonthDay{1975}{}{},
\newblock
\unskip
\newblock
\APACjournalVolNumPages{Cognition}{37}{2}{635}.
\PrintBackRefs{\CurrentBib}

\bibitem [\protect \citeauthoryear {%
Rutten%
}{%
Rutten%
}{%
{\protect \APACyear {2007}}%
}]{%
Rutten2007}
\APACinsertmetastar {%
Rutten2007}%
\begin{APACrefauthors}%
Rutten, R\BPBI J.%
\end{APACrefauthors}%
\unskip\
\newblock
\APACrefYearMonthDay{2007}{}{},
\newblock
{\BBOQ}\APACrefatitle {The Physics of Chromospheric Plasmas} {The Physics of
  Chromospheric Plasmas}.{\BBCQ}
\newblock
\BIn{} P.~Heinzel, I.~Dorotovic\BCBL {}\ \BBA {} R\BPBI J.~Rutten\ (\BEDS),
  \APACrefbtitle {ASP Conf. Ser.} {ASP Conf. Ser.}\ \BVOL~368, \BPG~27.
\PrintBackRefs{\CurrentBib}

\bibitem [\protect \citeauthoryear {%
Stix%
}{%
Stix%
}{%
{\protect \APACyear {2004}}%
}]{%
Stix2004}
\APACinsertmetastar {%
Stix2004}%
\begin{APACrefauthors}%
Stix, M.%
\end{APACrefauthors}%
\unskip\
\newblock
\APACrefYear{2004},
\newblock
\APACrefbtitle {Astronomy and Astrophysics Library} {Astronomy and Astrophysics
  Library}\ (\PrintOrdinal{2}\ \BEd).
\newblock
\APACaddressPublisher{Berlin}{Springer}.
\PrintBackRefs{\CurrentBib}

\bibitem [\protect \citeauthoryear {%
{Strunk Jr.}%
\ \BBA {} White%
}{%
{Strunk Jr.}%
\ \BBA {} White%
}{%
{\protect \APACyear {1979}}%
}]{%
Strunk1979}
\APACinsertmetastar {%
Strunk1979}%
\begin{APACrefauthors}%
{Strunk Jr.}, W.%
\BCBT {}\ \BBA {} White, E\BPBI B.%
\end{APACrefauthors}%
\unskip\
\newblock
\APACrefYear{1979},
\newblock
\APACrefbtitle {The Elements of Style} {The Elements of Style}\
  (\PrintOrdinal{3}\ \BEd).
\newblock
\APACaddressPublisher{New York}{MacMillan}.
\PrintBackRefs{\CurrentBib}

\end{thebibliography}


\begin{thebibliography}{}

\bibitem [\protect \citeauthoryear {%
{Colella}%
\ \BBA {} {Woodward}%
}{%
{Colella}%
\ \BBA {} {Woodward}%
}{%
{\protect \APACyear {1984}}%
}]{%
cw84}
\APACinsertmetastar {%
cw84}%
\begin{APACrefauthors}%
{Colella}, P.%
\BCBT {}\ \BBA {} {Woodward}, P\BPBI R.%
\end{APACrefauthors}%
\unskip\
\newblock
\APACrefYearMonthDay{1984}{{\APACmonth{09}}}{},
\newblock
\unskip
\newblock
\APACjournalVolNumPages{Journal of Computational Physics}{54}{}{174-201}.
\newblock
\begin{APACrefDOI} \doi{10.1016/0021-9991(84)90143-8} \end{APACrefDOI}
\PrintBackRefs{\CurrentBib}

\bibitem [\protect \citeauthoryear {%
{Frank}%
, {Dwarkadas}%
, {Panfichi}%
, {Crum}%
\BCBL {}\ \BBA {} {Burrows}%
}{%
{Frank}%
\ \protect \BOthers {.}}{%
{\protect \APACyear {2019}}%
}]{%
franketal19}
\APACinsertmetastar {%
franketal19}%
\begin{APACrefauthors}%
{Frank}, K\BPBI A.%
, {Dwarkadas}, V.%
, {Panfichi}, A.%
, {Crum}, R\BPBI M.%
\BCBL {}\ \BBA {} {Burrows}, D\BPBI N.%
\end{APACrefauthors}%
\unskip\
\newblock
\APACrefYearMonthDay{2019}{{\APACmonth{04}}}{},
\newblock
\unskip
\newblock
\APACjournalVolNumPages{\apj}{875}{}{14}.
\newblock
\begin{APACrefDOI} \doi{10.3847/1538-4357/ab0e81} \end{APACrefDOI}
\PrintBackRefs{\CurrentBib}

\bibitem [\protect \citeauthoryear {%
{Ghavamian}%
, {Rakowski}%
, {Hughes}%
\BCBL {}\ \BBA {} {Williams}%
}{%
{Ghavamian}%
\ \protect \BOthers {.}}{%
{\protect \APACyear {2003}}%
}]{%
ghavamianetal03}
\APACinsertmetastar {%
ghavamianetal03}%
\begin{APACrefauthors}%
{Ghavamian}, P.%
, {Rakowski}, C\BPBI E.%
, {Hughes}, J\BPBI P.%
\BCBL {}\ \BBA {} {Williams}, T\BPBI B.%
\end{APACrefauthors}%
\unskip\
\newblock
\APACrefYearMonthDay{2003}{Jun}{},
\newblock
\unskip
\newblock
\APACjournalVolNumPages{\apj}{590}{2}{833-845}.
\newblock
\begin{APACrefDOI} \doi{10.1086/375161} \end{APACrefDOI}
\PrintBackRefs{\CurrentBib}

\bibitem [\protect \citeauthoryear {%
{Hughes}%
, {Ghavamian}%
, {Rakowski}%
\BCBL {}\ \BBA {} {Slane}%
}{%
{Hughes}%
\ \protect \BOthers {.}}{%
{\protect \APACyear {2003}}%
}]{%
hughesetal03}
\APACinsertmetastar {%
hughesetal03}%
\begin{APACrefauthors}%
{Hughes}, J\BPBI P.%
, {Ghavamian}, P.%
, {Rakowski}, C\BPBI E.%
\BCBL {}\ \BBA {} {Slane}, P\BPBI O.%
\end{APACrefauthors}%
\unskip\
\newblock
\APACrefYearMonthDay{2003}{Jan}{},
\newblock
\unskip
\newblock
\APACjournalVolNumPages{\apjl}{582}{2}{L95-L99}.
\newblock
\begin{APACrefDOI} \doi{10.1086/367760} \end{APACrefDOI}
\PrintBackRefs{\CurrentBib}

\bibitem [\protect \citeauthoryear {%
{Leahy}%
}{%
{Leahy}%
}{%
{\protect \APACyear {2017}}%
}]{%
Leahyetal17}
\APACinsertmetastar {%
Leahyetal17}%
\begin{APACrefauthors}%
{Leahy}, D\BPBI A.%
\end{APACrefauthors}%
\unskip\
\newblock
\APACrefYearMonthDay{2017}{Mar}{},
\newblock
\unskip
\newblock
\APACjournalVolNumPages{\apj}{837}{1}{36}.
\newblock
\begin{APACrefDOI} \doi{10.3847/1538-4357/aa60c1} \end{APACrefDOI}
\PrintBackRefs{\CurrentBib}

\bibitem [\protect \citeauthoryear {%
{Leung}%
\ \BBA {} {Nomoto}%
}{%
{Leung}%
\ \BBA {} {Nomoto}%
}{%
{\protect \APACyear {2018}}%
}]{%
ln18}
\APACinsertmetastar {%
ln18}%
\begin{APACrefauthors}%
{Leung}, S\BHBI C.%
\BCBT {}\ \BBA {} {Nomoto}, K.%
\end{APACrefauthors}%
\unskip\
\newblock
\APACrefYearMonthDay{2018}{Jul}{},
\newblock
\unskip
\newblock
\APACjournalVolNumPages{\apj}{861}{2}{143}.
\newblock
\begin{APACrefDOI} \doi{10.3847/1538-4357/aac2df} \end{APACrefDOI}
\PrintBackRefs{\CurrentBib}

\bibitem [\protect \citeauthoryear {%
{Maeda}%
\ \BBA {} {Nomoto}%
}{%
{Maeda}%
\ \BBA {} {Nomoto}%
}{%
{\protect \APACyear {2003}}%
}]{%
mn03}
\APACinsertmetastar {%
mn03}%
\begin{APACrefauthors}%
{Maeda}, K.%
\BCBT {}\ \BBA {} {Nomoto}, K.%
\end{APACrefauthors}%
\unskip\
\newblock
\APACrefYearMonthDay{2003}{Dec}{},
\newblock
\unskip
\newblock
\APACjournalVolNumPages{\apj}{598}{2}{1163-1200}.
\newblock
\begin{APACrefDOI} \doi{10.1086/378948} \end{APACrefDOI}
\PrintBackRefs{\CurrentBib}

\bibitem [\protect \citeauthoryear {%
{Maeda}%
\ \protect \BOthers {.}}{%
{Maeda}%
\ \protect \BOthers {.}}{%
{\protect \APACyear {2010}}%
}]{%
maedaetal10}
\APACinsertmetastar {%
maedaetal10}%
\begin{APACrefauthors}%
{Maeda}, K.%
, {R{\"o}pke}, F\BPBI K.%
, {Fink}, M.%
, {Hillebrandt}, W.%
, {Travaglio}, C.%
\BCBL {}\ \BBA {} {Thielemann}, F\BPBI K.%
\end{APACrefauthors}%
\unskip\
\newblock
\APACrefYearMonthDay{2010}{Mar}{},
\newblock
\unskip
\newblock
\APACjournalVolNumPages{\apj}{712}{1}{624-638}.
\newblock
\begin{APACrefDOI} \doi{10.1088/0004-637X/712/1/624} \end{APACrefDOI}
\PrintBackRefs{\CurrentBib}

\bibitem [\protect \citeauthoryear {%
{Nomoto}%
, {Tominaga}%
, {Umeda}%
, {Kobayashi}%
\BCBL {}\ \BBA {} {Maeda}%
}{%
{Nomoto}%
\ \protect \BOthers {.}}{%
{\protect \APACyear {2006}}%
}]{%
nomotoetal06}
\APACinsertmetastar {%
nomotoetal06}%
\begin{APACrefauthors}%
{Nomoto}, K.%
, {Tominaga}, N.%
, {Umeda}, H.%
, {Kobayashi}, C.%
\BCBL {}\ \BBA {} {Maeda}, K.%
\end{APACrefauthors}%
\unskip\
\newblock
\APACrefYearMonthDay{2006}{Oct}{},
\newblock
\unskip
\newblock
\APACjournalVolNumPages{\nphysa}{777}{}{424-458}.
\newblock
\begin{APACrefDOI} \doi{10.1016/j.nuclphysa.2006.05.008} \end{APACrefDOI}
\PrintBackRefs{\CurrentBib}

\bibitem [\protect \citeauthoryear {%
{Peterson}%
, {Marshall}%
\BCBL {}\ \BBA {} {Andersson}%
}{%
{Peterson}%
\ \protect \BOthers {.}}{%
{\protect \APACyear {2007}}%
}]{%
spi}
\APACinsertmetastar {%
spi}%
\begin{APACrefauthors}%
{Peterson}, J\BPBI R.%
, {Marshall}, P\BPBI J.%
\BCBL {}\ \BBA {} {Andersson}, K.%
\end{APACrefauthors}%
\unskip\
\newblock
\APACrefYearMonthDay{2007}{Jan}{},
\newblock
\unskip
\newblock
\APACjournalVolNumPages{\apj}{655}{1}{109-127}.
\newblock
\begin{APACrefDOI} \doi{10.1086/509095} \end{APACrefDOI}
\PrintBackRefs{\CurrentBib}

\bibitem [\protect \citeauthoryear {%
{Rakowski}%
, {Ghavamian}%
\BCBL {}\ \BBA {} {Hughes}%
}{%
{Rakowski}%
\ \protect \BOthers {.}}{%
{\protect \APACyear {2003}}%
}]{%
rakowskietal03}
\APACinsertmetastar {%
rakowskietal03}%
\begin{APACrefauthors}%
{Rakowski}, C\BPBI E.%
, {Ghavamian}, P.%
\BCBL {}\ \BBA {} {Hughes}, J\BPBI P.%
\end{APACrefauthors}%
\unskip\
\newblock
\APACrefYearMonthDay{2003}{Jun}{},
\newblock
\unskip
\newblock
\APACjournalVolNumPages{\apj}{590}{2}{846-857}.
\newblock
\begin{APACrefDOI} \doi{10.1086/375162} \end{APACrefDOI}
\PrintBackRefs{\CurrentBib}

\bibitem [\protect \citeauthoryear {%
{Rakowski}%
, {Ghavamian}%
\BCBL {}\ \BBA {} {Laming}%
}{%
{Rakowski}%
\ \protect \BOthers {.}}{%
{\protect \APACyear {2009}}%
}]{%
R09}
\APACinsertmetastar {%
R09}%
\begin{APACrefauthors}%
{Rakowski}, C\BPBI E.%
, {Ghavamian}, P.%
\BCBL {}\ \BBA {} {Laming}, J\BPBI M.%
\end{APACrefauthors}%
\unskip\
\newblock
\APACrefYearMonthDay{2009}{May}{},
\newblock
\unskip
\newblock
\APACjournalVolNumPages{\apj}{696}{2}{2195-2205}.
\newblock
\begin{APACrefDOI} \doi{10.1088/0004-637X/696/2/2195} \end{APACrefDOI}
\PrintBackRefs{\CurrentBib}

\bibitem [\protect \citeauthoryear {%
{Russell}%
\ \BBA {} {Dopita}%
}{%
{Russell}%
\ \BBA {} {Dopita}%
}{%
{\protect \APACyear {1992}}%
}]{%
rd92}
\APACinsertmetastar {%
rd92}%
\begin{APACrefauthors}%
{Russell}, S\BPBI C.%
\BCBT {}\ \BBA {} {Dopita}, M\BPBI A.%
\end{APACrefauthors}%
\unskip\
\newblock
\APACrefYearMonthDay{1992}{Jan}{},
\newblock
\unskip
\newblock
\APACjournalVolNumPages{\apj}{384}{}{508}.
\newblock
\begin{APACrefDOI} \doi{10.1086/170893} \end{APACrefDOI}
\PrintBackRefs{\CurrentBib}

\bibitem [\protect \citeauthoryear {%
{Seitenzahl}%
\ \protect \BOthers {.}}{%
{Seitenzahl}%
\ \protect \BOthers {.}}{%
{\protect \APACyear {2013}}%
}]{%
seitenzahletal13}
\APACinsertmetastar {%
seitenzahletal13}%
\begin{APACrefauthors}%
{Seitenzahl}, I\BPBI R.%
, {Ciaraldi-Schoolmann}, F.%
, {R{\"o}pke}, F\BPBI K.%
\ et al.\end{APACrefauthors}%
\unskip\
\newblock
\APACrefYearMonthDay{2013}{Feb}{},
\newblock
\unskip
\newblock
\APACjournalVolNumPages{\mnras}{429}{2}{1156-1172}.
\newblock
\begin{APACrefDOI} \doi{10.1093/mnras/sts402} \end{APACrefDOI}
\PrintBackRefs{\CurrentBib}

\bibitem [\protect \citeauthoryear {%
{Sukhbold}%
, {Ertl}%
, {Woosley}%
, {Brown}%
\BCBL {}\ \BBA {} {Janka}%
}{%
{Sukhbold}%
\ \protect \BOthers {.}}{%
{\protect \APACyear {2016}}%
}]{%
sukhboldetal16}
\APACinsertmetastar {%
sukhboldetal16}%
\begin{APACrefauthors}%
{Sukhbold}, T.%
, {Ertl}, T.%
, {Woosley}, S\BPBI E.%
, {Brown}, J\BPBI M.%
\BCBL {}\ \BBA {} {Janka}, H\BPBI T.%
\end{APACrefauthors}%
\unskip\
\newblock
\APACrefYearMonthDay{2016}{Apr}{},
\newblock
\unskip
\newblock
\APACjournalVolNumPages{\apj}{821}{1}{38}.
\newblock
\begin{APACrefDOI} \doi{10.3847/0004-637X/821/1/38} \end{APACrefDOI}
\PrintBackRefs{\CurrentBib}

\bibitem [\protect \citeauthoryear {%
{van der Heyden}%
, {Bleeker}%
, {Kaastra}%
\BCBL {}\ \BBA {} {Vink}%
}{%
{van der Heyden}%
\ \protect \BOthers {.}}{%
{\protect \APACyear {2003}}%
}]{%
vanderheydenetal03}
\APACinsertmetastar {%
vanderheydenetal03}%
\begin{APACrefauthors}%
{van der Heyden}, K\BPBI J.%
, {Bleeker}, J\BPBI A\BPBI M.%
, {Kaastra}, J\BPBI S.%
\BCBL {}\ \BBA {} {Vink}, J.%
\end{APACrefauthors}%
\unskip\
\newblock
\APACrefYearMonthDay{2003}{{\APACmonth{07}}}{},
\newblock
\unskip
\newblock
\APACjournalVolNumPages{\aap}{406}{}{141-148}.
\newblock
\begin{APACrefDOI} \doi{10.1051/0004-6361:20030658} \end{APACrefDOI}
\PrintBackRefs{\CurrentBib}

\end{thebibliography}

\end{document}